\documentclass[12pt]{article}
\usepackage{amssymb}
\begin{document}



\def\a{\alpha}
\def\b{\beta}
\def\d{\delta}
\def\e{\epsilon}
\def\g{\gamma}
\def\h{\mathfrak{h}}
\def\k{\kappa}
\def\l{\lambda}
\def\o{\omega}
\def\p{\wp}
\def\r{\rho}
\def\t{\theta}
\def\s{\sigma}
\def\z{\zeta}
\def\x{\xi}
 \def\A{{\cal{A}}}
 \def\B{{\cal{B}}}
 \def\C{{\cal{C}}}
 \def\D{{\cal{D}}}
\def\G{\Gamma}
\def\K{{\cal{K}}}
\def\O{\Omega}
\def\L{\Lambda}
\def\f{E_{\tau,\eta}(sl_2)}
\def\E{E_{\tau,\eta}(sl_n)}
\def\Zb{\mathbf{Z}}
\def\Cb{\mathbb{C}}

\def\R{\overline{R}}

\def\beq{\begin{equation}}
\def\eeq{\end{equation}}
\def\bea{\begin{eqnarray}}
\def\eea{\end{eqnarray}}
\def\ba{\begin{array}}
\def\ea{\end{array}}
\def\no{\nonumber}
\def\le{\langle}
\def\re{\rangle}
\def\lt{\left}
\def\rt{\right}

\newtheorem{Theorem}{Theorem}
\newtheorem{Definition}{Definition}
\newtheorem{Proposition}{Proposition}
\newtheorem{Lemma}{Lemma}
\newtheorem{Corollary}{Corollary}
\newcommand{\proof}[1]{{\bf Proof. }
        #1\begin{flushright}$\Box$\end{flushright}}

\baselineskip=20pt

\newfont{\elevenmib}{cmmib10 scaled\magstep1}
\newcommand{\preprint}{
   \begin{flushright}\normalsize  \sf
     {\tt hep-th/0504048} \\ April 2005
   \end{flushright}}
\newcommand{\Title}[1]{{\baselineskip=26pt
   \begin{center} \Large \bf #1 \\ \ \\ \end{center}}}
\newcommand{\Author}{\begin{center}
   \large \bf
Wen-Li Yang${}^{a,b}$
 ~ and~Yao-Zhong Zhang${}^b$\end{center}}
\newcommand{\Address}{\begin{center}

     ${}^a$ Institute of Modern Physics, Northwest University
     Xian 710069, P.R. China\\
     ${}^b$ Department of Mathematics, The University of Queensland,
     Brisbane 4072, Australia
   \end{center}}
\newcommand{\Accepted}[1]{\begin{center}
   {\large \sf #1}\\ \vspace{1mm}{\small \sf Accepted for Publication}
   \end{center}}

\preprint
\thispagestyle{empty}
\bigskip\bigskip\bigskip

\Title{A note on the non-diagonal K-matrices for the trigonometric
$A^{(1)}_{n-1}$ vertex  model } \Author

\Address
\vspace{1cm}

\begin{abstract}
This note presents  explicit matrix expressions of a class of
recently-discovered non-diagonal K-matrices for the trigonometric
$A^{(1)}_{n-1}$ vertex model. From these explicit expressions, it
is easily seen   that in addition to a {\it discrete\/} (positive
integer) parameter $l$, $1\leq l\leq n$, the K-matrices contain
$n+1$ (or $n$) continuous free boundary parameters.

\vspace{1truecm}

\end{abstract}

\vskip 1.0truecm

Inspired from the observation \cite{Yan043} that the generic
non-diagonal solutions (or K-matrices) \cite{Dev93,Gho94} of the
reflection equation for the spin-$\frac{1}{2}$ XXZ model are
decomposed into the product of intertwiner-matrices and  diagonal
face-type K-matrix, in \cite{Yan04-N} an intertwiner-matrix
approach was developed and used to construct a class of
non-diagonal solutions of the reflection equation for the
trigonometric $A^{(1)}_{n-1}$ vertex model. There the K-matrices
were  expressed in terms of the intertwiner-matrix and a diagonal
matrix. To fully realize the application of the solutions obtained
in \cite{Yan04-N}, it may be useful to write them  in {\it
explicit\/} and {\it familiar\/} matrix form. The purpose of this
note is to provide such explicit expressions. From these
expressions it is easily seen that  in addition to a {\it
discrete\/} (positive integer) parameter $l$, $1\leq l\leq n$, the
solutions we constructed in \cite{Yan04-N} contain $n+1$ (or $n$)
continuous free boundary parameters and have $3n-2$ (or $2n-1$)
non-vanishing matrix elements.

Our starting point in \cite{Yan04-N} is the trigonometric R-matrix
associated with the $n$-dimensional representation of
$A^{(1)}_{n-1}$ given in  \cite{Per81,Baz91}: \bea
\hspace{-0.5cm}R(u)=\sum_{\a=1}^{n}R^{\a\a}_{\a\a}(u)E_{\a\a}\otimes
E_{\a\a} +\sum_{\a\ne \b}\lt\{R^{\a\b}_{\a\b}(u)E_{\a\a}\otimes
E_{\b\b}+ R^{\b\a}_{\a\b}(u)E_{\b\a}\otimes
E_{\a\b}\rt\},\label{R-matrix} \eea where $E_{ij}$ is the matrix
with elements $(E_{ij})^l_k=\d_{jk}\d_{il}$. The coefficient
functions are \bea
R^{\a\b}_{\a\b}(u)&=&\lt\{\begin{array}{cc}\frac{\sin(u)\,e^{-i\eta}}{\sin(u+\eta)},&
\a>\b,\\[6pt]1,&\a=\b,\\[6pt]\frac{\sin(u)\,e^{i\eta}}{\sin(u+\eta)},&
\a<\b,\end{array}\rt.\label{Elements1}\\[6pt]
R^{\b\a}_{\a\b}(u)&=&\lt\{\begin{array}{cc}\frac{\sin(\eta)\,e^{iu}}{\sin(u+\eta)},&
\a>\b,\\[6pt]1,&\a=\b,\\[6pt]\frac{\sin(\eta)\,e^{-iu}}{\sin(u+\eta)},&
\a<\b.\end{array}\rt.\label{Elements2}\eea Here $\eta$ is the
so-called crossing parameter. In addition to the quantum
Yang-Baxter equation, the R-matrix satisfies the following
unitarity, crossing-unitarity and quasi-classical
relations:\begin{eqnarray}
 &&\hspace{-1.5cm}\mbox{
 Unitarity}:\hspace{42.5mm}R_{12}(u)R_{21}(-u)= {\rm id},\label{Unitarity}\\
 &&\hspace{-1.5cm}\mbox{
 Crossing-unitarity}:\quad
 R^{t_2}_{12}(u)M_2^{-1}R_{21}^{t_2}(-u-n\eta)M_2
 = \frac{\sin(u)\sin(u+n\eta)}{\sin(u+\eta)\sin(u+n\eta-\eta)}\,\mbox{id},
 \label{crosing-unitarity}\\
 &&\hspace{-1.5cm}\mbox{ Quasi-classical
 property}:\hspace{22.5mm}\, R_{12}(u)|_{\eta\rightarrow 0}= {\rm
id}.\label{quasi}
\end{eqnarray}
Here $R_{21}(u)=P_{12}R_{12}(u)P_{12}$ with $P_{12}$ being the
usual permutation operator and $t_i$ denotes the transposition in
the $i$-th space. The crossing matrix $M$ is a diagonal $n\times
n$ matrix with elements \bea
M_{\a\b}=M_{\a}\d_{\a\b},~~M_{\a}=e^{-2i\a\eta},~\a=1,\cdots,n.\label{C-Matrix}\eea

Boundary K-matrices $K^-(u)$ and $K^{+}(u)$, which give rise to
integrable boundary conditions of an open chain on the right and
left boundaries, respectively, satisfy the reflection and dual
reflection equations \cite{Skl88,Mez91}:
 \begin{eqnarray}
 &&R_{12}(u_1-u_2)K^-_1(u_1)R_{21}(u_1+u_2)K^-_2(u_2)\no\\
  &&\qquad\quad=
 K^-_2(u_2)R_{12}(u_1+u_2)K^-_1(u_1)R_{21}(u_1-u_2),\label{RE-V}\\
&&R_{12}(u_2-u_1)K^+_1(u_1)\,M_1^{-1}\,R_{21}(-u_1-u_2-n\eta)\,M_1\,K^+_2(u_2)\no\\
&&\qquad\quad=
M_1\,K^+_2(u_2)R_{12}(-u_1-u_2-n\eta)\,M_1^{-1}\,K^{+}_1(u_1)R_{21}(u_2-u_1).
\label{DRE-V1}\eea Different  integrable boundary conditions are
described by different solutions $K^{-}(u)$ ($K^{+}(u)$) to the
(dual) reflection equation \cite{Skl88, Gho94}.

Let us briefly recall some of the results in \cite{Yan04-N}. Let
$\lt\{\e_{i}~|~i=1,2,\cdots,n\rt\}$ be the orthonormal basis of
the vector space $\Cb^n$ such that $\langle\e_i,~\e_j
\rangle=\d_{ij}$. For a generic vector $\l\in \Cb^n$, define \bea
\l_i=\langle \l,\e_i\rangle,
~~|\l|=\sum_{k=1}^n\l_k,~~i=1,\cdots,n. \label{Def1}\eea Let us
introduce an $n\times n$ matrix $\Phi(u;\l)$ which depends on the
spectrum parameter $u$ and $\l$. The non-vanishing matrix elements
of $\Phi(u;\l)$ are given by  \bea
\lt(\begin{array}{ccccccc}e^{i\eta f_1(\l)} &&&&&&e^{i\eta
F_n(\l)+\rho}e^{2iu}\\e^{i\eta F_1(\l)}&e^{i\eta f_2(\l)}&&&&&\\
&e^{i\eta F_2(\l)}&\ddots&&&&\\&&\ddots&e^{i\eta
f_j(\l)}&&&\\&&&e^{i\eta F_j(\l)}&\ddots&&\\&&&&\ddots&e^{i\eta
f_{n-1}(\l)}&\\&&&&&e^{i\eta F_{n-1}(\l)}&e^{i\eta f_n(\l)}
\end{array}\rt).\label{In-matrix}\eea
Here $\rho$  is a complex constant with regard to $u$ and $\l$,
and $\{f_i(\l)|i=1,\ldots,n\}$ and $\{F_{i}(\l)|i=1,\ldots,n\}$
are linear functions of $\l$:\bea
f_i(\l)&=&\sum_{k=1}^{i-1}\l_k-\l_i-\frac{1}{2}|\l|,~~i=1,\ldots,n,\label{function1}\\
F_{i}(\l)&=&\sum_{k=1}^{i}\l_k-\frac{1}{2}|\l|,~~i=1,\ldots,n-1,\label{function2}\\
F_n(\l)&=&-\frac{3}{2}|\l|.\label{function3}\eea The determinant
of $\Phi(u;\l)$ is \cite{Yan04-N}\bea {\rm
Det}\lt(\Phi(u;\l)\rt)=e^{i\eta\sum_{k=1}^{n}\frac{n-2(k+1)}{2}\l_k}\,
(1-(-1)^ne^{2iu+\rho}).\label{Det}\eea For a generic $\rho\in\Cb$
this determinant is not vanishing and thus the inverse of
$\Phi(u;\l)$  exists. Associated to a positive integer $l$ $(1\leq
l\leq n)$, let us introduce a diagonal matrix \bea D^{(l)}(u)={\rm
Diag} (k^{(l)}_1(u),\ldots,k^{(l)}_n(u)),\eea where
$\{k^{(l)}_i(u)|i=1,\ldots,n\}$ are \bea k^{(l)}_j(u)
&=&\lt\{\begin{array}{ll}1,&1\leq j\leq l,\\[4pt]
\frac{\sin(\xi-u)}{\sin(\xi+u)}e^{-2iu},&l+1\leq j\leq
n.\end{array}\rt.\label{K-matrix1}\eea Here $\xi$ is free complex
parameter. Then one can define the non-diagonal K-matrices
$\{K^{(l)}(u)|l=1,\ldots,n\}$ associated with
$\{D^{(l)}(u)|l=1,\ldots,n\}$ and $\Phi(u;\l)$ as follows
\cite{Yan04-N}:\bea
K^{(l)}(u)=\Phi(u;\l)D^{(l)}(u)\lt\{\Phi(-u;\l)\rt\}^{-1},
~~l=1,\ldots,n.\label{K-m}\eea It has been shown in \cite{Yan04-N}
that the matrix $\Phi(u;\l)$ given by (\ref{In-matrix}) is the
intertwiner-matrix which intertwines two trigonometric R-matrices,
and thus  the non-diagonal K-matrices $\{K^{(l)}(u)\}$ given by
(\ref{K-m}) solve the reflection equation (\ref{RE-V}) for the
trigonometric $A^{(1)}_{n-1}$ vertex model. Moreover, (\ref{K-m})
implies that the K-matrices satisfy the regular condition $
K^{(l)}(0)={\rm id},~l=1,\ldots,n,$ and boundary unitarity
relation $K^{(l)}(u)K^{(l)}(-u)={\rm id}, ~l=1,\ldots,n.$

Through a tedious calculation for $n=2,3,4,5$ with the help of
Mathematica program, we reconfirm the following properties for the
non-diagonal K-matrices (\ref{K-m}): $K^{(l)}(u)$
$(l=1,\ldots,n-1)$ depend on $n+1$ continuous free parameters
$\xi$, $\{\l_i|i=1,\ldots,n-1\}$  and $\rho$, and have $3n-2$
non-vanishing matrix elements (c.f. \cite{Lim02,Mal04});
$K^{(n)}(u)$ depends on $n$ continuous free parameters
$\{\l_i|i=1,\ldots,n-1\}$ and $\rho$, and has $2n-1$ non-vanishing
matrix elements (c.f. \cite{Lim02,Mal04}). The dependence on
$\l_n$ {\it disappears\/} in the final expressions of the
K-matrices although it appears in the expression of $\Phi(u;\l)$.
The above properties are expected to  hold for generic $n$. In the
rational limit, the trigonometric K-matrices (\ref{K-m}) reduce to
those corresponding to  the rational $A^{(1)}_{n-1}$ vertex model
\cite{Min01,Gal04} with a special choice of the
spectral-independent similarity transformation matrix.

In the following, we  give the explicit matrix expressions of the
K-matrices (\ref{K-m}) for the cases $n=3,4$.

\vskip0.3in

 \noindent {\large \bf The $A^{(1)}_2$ case:}
\vskip0.2in

 \noindent There are three types of K-matrices for the
 trigonometric $A^{(1)}_{2}$ model.
\begin{itemize}
\item For the K-matrix $K^{(1)}(u)$, the $7$ non-vanishing matrix
elements $K(u)^k_j$  are given by: \bea
&&K(u)^1_1=\frac{e^{2iu}}{e^{2iu}+e^{\rho}}
\lt(1-e^{\rho}\frac{\sin(u-\xi)}{\sin(u+\xi)}\rt),\,\,
K(u)^1_2=\frac{e^{-2i\eta\l_1+\rho}}{e^{2iu}+e^{\rho}}
\lt(1+e^{2iu}\frac{\sin(u-\xi)}{\sin(u+\xi)}\rt),\no\\
&&K(u)^1_3=-\frac{e^{-2i\eta(\l_1+\l_2)+\rho}}{e^{2iu}+e^{\rho}}
\lt(1+e^{2iu}\frac{\sin(u-\xi)}{\sin(u+\xi)}\rt),\no\\
&&K(u)^2_1=\frac{e^{2i\eta\l_1+i(u+\xi)}\sin2u}
{\lt(e^{2iu}+e^{\rho}\rt)\sin(u+\xi)},\,\,
K(u)^2_2=\frac{1}{e^{2iu}+e^{\rho}}
\lt(e^{\rho}-\frac{\sin(u-\xi)}{\sin(u+\xi)}\rt),\no\\
&&K(u)^2_3=-\frac{e^{-2i\eta\l_2-i(u-\xi)+\rho}\sin2u}{\lt(e^{2iu}+e^{\rho}\rt)
\sin(u+\xi)},\,\,
K(u)^3_3=e^{-2iu}\frac{\sin(\xi-u)}{\sin(\xi+u)}.\label{K1-1} \eea

\item For the K-matrix $K^{(2)}(u)$, the $7$ non-vanishing matrix
elements $K(u)^k_j$  are given by: \bea
&&K(u)^1_1=\frac{e^{2iu}}{e^{2iu}+e^{\rho}}
\lt(1-e^{\rho}\frac{\sin(u-\xi)}{\sin(u+\xi)}\rt),\,\,
K(u)^1_2=\frac{e^{-2i\eta\l_1+\rho}}{e^{2iu}+e^{\rho}}
\lt(1+e^{2iu}\frac{\sin(u-\xi)}{\sin(u+\xi)}\rt),\no\\
&&K(u)^1_3=-\frac{e^{-2i\eta(\l_1+\l_2)+\rho}}{e^{2iu}+e^{\rho}}
\lt(1+e^{2iu}\frac{\sin(u-\xi)}{\sin(u+\xi)}\rt),\no\\
&&K(u)^2_2=1,\no\\
&&K(u)^3_1=-\frac{e^{2i\eta(\l_1+\l_2)+i(u+\xi)}\sin2u}
{\lt(e^{2iu}+e^{\rho}\rt)\sin(u+\xi)},\,\,
K(u)^3_2=\frac{e^{2i\eta\l_2+i(u+\xi)}\sin2u}{\lt(e^{2iu}+e^{\rho}\rt)
\sin(u+\xi)},\no\\
&&K(u)^3_3=\frac{1}{e^{2iu}+e^{\rho}}
\lt(e^{\rho}-\frac{\sin(u-\xi)}{\sin(u+\xi)}\rt). \eea

\item For the K-matrix $K^{(3)}(u)$, the $5$ non-vanishing matrix
elements $K(u)^k_j$  are given by: \bea
&&K(u)^1_1=\frac{e^{2iu}+e^{4iu+\rho}}{e^{2iu}+e^{\rho}},\,\,
K(u)^1_2=-\frac{e^{-2i\eta\l_1+\rho}\lt(e^{4iu}-1\rt)}{e^{2iu}+e^{\rho}},\no\\
&&K(u)^1_3=\frac{e^{-2i\eta(\l_1+\l_2)+\rho}\lt(e^{4iu}-1\rt)}{e^{2iu}+e^{\rho}},
\,\,K(u)^2_2=K(u)^3_3=1.\label{K1-2}\eea

\end{itemize}

\noindent {\large\bf The $A^{(1)}_3$ case:}

\vskip0.2in

\noindent There are four types of K-matrices for the trigonometric
$A^{(1)}_{3}$ model.
\begin{itemize}
\item For the K-matrix $K^{(1)}(u)$, the $10$ non-vanishing matrix
elements $K(u)^k_j$  are given by:

\bea &&K(u)^1_1=\frac{e^{2iu}}{e^{2iu}-e^{\rho}}
\lt(1+e^{\rho}\frac{\sin(u-\xi)}{\sin(u+\xi)}\rt),\,\,
K(u)^1_2=-\frac{e^{-2i\eta\l_1+\rho}}{e^{2iu}-e^{\rho}}
\lt(1+e^{2iu}\frac{\sin(u-\xi)}{\sin(u+\xi)}\rt),\no\\
&&K(u)^1_3=\frac{e^{-2i\eta(\l_1+\l_2)+\rho}}{e^{2iu}-e^{\rho}}
\lt(1+e^{2iu}\frac{\sin(u-\xi)}{\sin(u+\xi)}\rt),\,\,\no\\
&&K(u)^1_4=-\frac{e^{-2i\eta(\l_1+\l_2+\l_3)+\rho}}{e^{2iu}-e^{\rho}}
\lt(1+e^{2iu}\frac{\sin(u-\xi)}{\sin(u+\xi)}\rt),\,\,
K(u)^2_1=\frac{e^{2i\eta\l_1+i(u+\xi)}\sin2u}
{\lt(e^{2iu}-e^{\rho}\rt)\sin(u+\xi)},\no\\
&&K(u)^2_2=-\frac{1}{e^{2iu}-e^{\rho}}
\lt(e^{\rho}+\frac{\sin(u-\xi)}{\sin(u+\xi)}\rt),\,\,
K(u)^2_3=\frac{e^{-2i\eta\l_2-i(u-\xi)+\rho}\sin2u}{\lt(e^{2iu}-e^{\rho}\rt)
\sin(u+\xi)},\no\\
&&K(u)^2_4=-\frac{e^{-2i\eta(\l_2+\l_3)-i(u-\xi)+\rho}\sin2u}{\lt(e^{2iu}-e^{\rho}\rt)
\sin(u+\xi)},\,\,K(u)^3_3=K(u)^4_4=e^{-2iu}\frac{\sin(\xi-u)}{\sin(\xi+u)}.
\label{K2-1}\eea

\item For the K-matrix $K^{(2)}(u)$, the $10$ non-vanishing matrix
elements $K(u)^k_j$  are given by:

\bea &&K(u)^1_1=\frac{e^{2iu}}{e^{2iu}-e^{\rho}}
\lt(1+e^{\rho}\frac{\sin(u-\xi)}{\sin(u+\xi)}\rt),\,\,
K(u)^1_2=-\frac{e^{-2i\eta\l_1+\rho}}{e^{2iu}-e^{\rho}}
\lt(1+e^{2iu}\frac{\sin(u-\xi)}{\sin(u+\xi)}\rt),\no\\
&&K(u)^1_3=\frac{e^{-2i\eta(\l_1+\l_2)+\rho}}{e^{2iu}-e^{\rho}}
\lt(1+e^{2iu}\frac{\sin(u-\xi)}{\sin(u+\xi)}\rt),\,\,\no\\
&&K(u)^1_4=-\frac{e^{-2i\eta(\l_1+\l_2+\l_3)+\rho}}{e^{2iu}-e^{\rho}}
\lt(1+e^{2iu}\frac{\sin(u-\xi)}{\sin(u+\xi)}\rt),\,\,K(u)^2_2=1,\no\\
&&K(u)^3_1=-\frac{e^{2i\eta(\l_1+\l_2)+i(u+\xi)}\sin2u}
{\lt(e^{2iu}-e^{\rho}\rt)\sin(u+\xi)},\,\,
K(u)^3_2=\frac{e^{2i\eta(\l_2)+i(u+\xi)}\sin2u}
{\lt(e^{2iu}-e^{\rho}\rt)\sin(u+\xi)},\no\\
&&K(u)^3_3=-\frac{1}{e^{2iu}-e^{\rho}}
\lt(e^{\rho}+\frac{\sin(u-\xi)}{\sin(u+\xi)}\rt),\,\,
K(u)^3_4=\frac{e^{-2i\eta\l_3-i(u-\xi)+\rho}\sin2u}{\lt(e^{2iu}-e^{\rho}\rt)
\sin(u+\xi)},\no\\
&&K(u)^4_4=e^{-2iu}\frac{\sin(\xi-u)}{\sin(\xi+u)}. \eea

\item For the K-matrix $K^{(3)}(u)$, the $10$ non-vanishing matrix
elements $K(u)^k_j$  are given by:

\bea &&K(u)^1_1=\frac{e^{2iu}}{e^{2iu}-e^{\rho}}
\lt(1+e^{\rho}\frac{\sin(u-\xi)}{\sin(u+\xi)}\rt),\,\,
K(u)^1_2=-\frac{e^{-2i\eta\l_1+\rho}}{e^{2iu}-e^{\rho}}
\lt(1+e^{2iu}\frac{\sin(u-\xi)}{\sin(u+\xi)}\rt),\no\\
&&K(u)^1_3=\frac{e^{-2i\eta(\l_1+\l_2)+\rho}}{e^{2iu}-e^{\rho}}
\lt(1+e^{2iu}\frac{\sin(u-\xi)}{\sin(u+\xi)}\rt),\,\,\no\\
&&K(u)^1_4=-\frac{e^{-2i\eta(\l_1+\l_2+\l_3)+\rho}}{e^{2iu}-e^{\rho}}
\lt(1+e^{2iu}\frac{\sin(u-\xi)}{\sin(u+\xi)}\rt),\,\,K(u)^2_2=K(u)^3_3=1,\no\\
&&K(u)^4_1=\frac{e^{2i\eta(\l_1+\l_2+\l_3)+i(u+\xi)}\sin2u}
{\lt(e^{2iu}-e^{\rho}\rt)\sin(u+\xi)},\,\,
K(u)^4_2=-\frac{e^{2i\eta(\l_2+\l_3)+i(u+\xi)}\sin2u}
{\lt(e^{2iu}-e^{\rho}\rt)\sin(u+\xi)},\no\\
&&K(u)^4_3=\frac{e^{2i\eta\l_3+i(u+\xi)}\sin2u}{\lt(e^{2iu}-e^{\rho}\rt)
\sin(u+\xi)},\,\,K(u)^4_4=-\frac{1}{e^{2iu}-e^{\rho}}
\lt(e^{\rho}+\frac{\sin(u-\xi)}{\sin(u+\xi)}\rt).\eea

\item For the K-matrix $K^{(4)}(u)$, the $7$ non-vanishing matrix
elements $K(u)^k_j$  are given by: \bea
&&K(u)^1_1=\frac{e^{2iu}-e^{4iu+\rho}}{e^{2iu}-e^{\rho}},\,\,
K(u)^1_2=\frac{e^{-2i\eta\l_1+\rho}\lt(e^{4iu}-1\rt)}{e^{2iu}-e^{\rho}},\no\\
&&K(u)^1_3=-\frac{e^{-2i\eta(\l_1+\l_2)+\rho}\lt(e^{4iu}-1\rt)}{e^{2iu}-e^{\rho}},
\,\,K(u)^1_4=\frac{e^{-2i\eta(\l_1+\l_2+\l_3)+\rho}\lt(e^{4iu}-1\rt)}
{e^{2iu}-e^{\rho}},\no\\
&&K(u)^2_2=K(u)^3_3=K(u)^4_4=1.\label{K2-2}\eea

\end{itemize}

In summary, we have presented the explicit matrix expressions of
the non-diagonal K-matrices obtained in \cite{Yan04-N} for the
trigonometric $A^{(1)}_{n-1}$ vertex model. From these results, it
is easily seen that the K-matrices $K^{(l)}(u)\,(l=1,\ldots,n-1)$
depend on $n+1$ continuous free parameters and have $3n-2$
non-vanishing matrix elements, and that the K-matrix $K^{(n)}(u)$
depends on $n$ continuous free parameters and has $2n-1$
non-vanishing matrix elements.

This work was financially supported by the Australian Research
Council.


\end{document}